
%
\newdimen\bsk \bsk=12pt \newdimen\pbg \pbg=4pt \newdimen\mbg \mbg=4pt
\def\varvskip#1{\vskip #1\bsk plus #1\pbg minus #1\mbg}
\font\Bigbf=cmbx10 scaled \magstep2
\font\bigbf=cmbx10 scaled \magstep1
\def\section#1{{\noindent{\bigbf #1}\par}}
\def\eq#1{eq.~(#1)}
\magnification=\magstep1
\parskip=0.5\baselineskip
%
\def\St{Stokes' theorem}
\let\ep=\epsilon
\let\tl=\tilde
\let\th=\theta
\def\d{\partial}
\def\dd{{\rm d}}
\def\G{\Gamma}
\def\Gd{G_{\!D}}
\def\B{{\cal B}}
\def\M{{\cal M}}
\def\S{{\cal S}}
%
\def\IJMP#1{{\sl Int.\ J.\ Mod.\ Phys.\ }{\bf #1}}
\def\NP#1{{\sl Nucl.\ Phys.\ }{\bf #1}}
\def\PR#1{{\sl Phys.\ Rev.\ }{\bf #1}}
\def\PRp#1{{\sl Phys. Rep.\ }{\bf #1}}
\def\cite#1{[{#1}]}
\def\ref#1#2{\item{[#1]} #2}
\let\RevLoops             =1
\let\RevGrav              =2
\let\RevTFT               =3
\let\RevTurb              =4
\let\Bralic               =5
\let\OtherNonAbelianStokes=6
\let\KalbRamondNambu      =7
\hbox{}
\varvskip{6}
\centerline{\Bigbf
            A generalized ``surfaceless'' Stokes' theorem
}
\varvskip{6}
\centerline{\bigbf
                             N. Brali\'c
}
\varvskip{2}
\centerline{
{\it Facultad de F\'\i{}sica, Pontificia Universidad Cat\'olica de Chile}
}
\centerline{
{\it Casilla 306, Santiago 22, Chile. E-mail:\/} {\tt nbralic@lascar.puc.cl}
}
\varvskip{6}
{\narrower\noindent
We derive a generalized \St, valid in any dimension and for arbitrary
loops, even if self intersecting or knotted.  The generalized theorem
does not involve an auxiliary surface, but inherits a higher rank
gauge symmetry from the invariance under deformations of the surface
used in the conventional formulation.
\par}
\vfil
hep-th/yymmnn
\eject
\section{1. Introduction}
\noindent
\St{} can be stated in general terms as
$$
\G[\d\S] = \oint\limits_{\d\S} A = \int\limits_\S \dd A ,
\eqno(1.1)
$$
where $\S$ is a compact, orientable, $(n+1)$-dimensional manifold
with boundary $\d\S$, $A$ is an $n$-form field on $\S$ and $\dd A$
is its exterior derivative.  We shall soon scale down this
high-browed language of forms to that of everyday tensor analysis.
But before we do that let us recall some (very) well known facts.

First, it is important to note that $\G[\d\S]$ in \eq{1.1} depends
on the boundary $\d\S$ of $\S$, but not on $\S$ itself.  If $\S$
and $\S'$ have the same boundary $\d\S$, they define a new closed
orientable manifold $\S - \S'$, so we can use \St{} again to show
that the right hand sides of \eq{1.1}, computed with $\S$ and $\S'$,
differ by
$$
\oint\limits_{\S-\S'} \dd A  = \int \dd\,\dd A = 0 .
\eqno(1.2)
$$
Secondly, $\G[\d\S]$ is invariant under the (generalized) Abelian
gauge transformations given by
$$
A \to A' = A + \dd\Lambda ,
\eqno(1.3)
$$
where $\Lambda$ is an $(n-1)$-form field.  Under this transformations
$\dd A$ itself is invariant
$$
\dd A' = \dd A ,
$$
so we can think of the invariance of $\G[\d\S]$ as following from
the right hand side of \eq{1.1}.  But we can also think of the
invariance of $\G[\d\S]$ as a consequence of $\d\S$ being closed:
using \St{} once more, the change of $\G[\d\S]$ under the
transformation in \eq{1.3} would be
$$
\G' - \G = \oint\limits_{\d\S} \dd\Lambda
         = \oint\limits_{\d\,\d\S} \Lambda = 0 .
$$

The most important physical applications are with $n=1$, in which
case \St{} relates the line integral of a covariant vector field
$A$ along a closed curve $q$, to the flux of the curl $F$ of $A$
through a (compact, orientable) surface $\S$ with boundary $\d\S=q$.
As is often the case, we shall think of $q$ and $\S$ as embedded
in a $D$-dimensional manifold $\M$.  Then, using old-fashioned
indices, \St{} is
$$
\G[q] = \oint\limits_{q} dx^{\mu} A_\mu(x)
      = \int\limits_{\S} d^2\!\sigma^{\mu\nu} F_{\mu\nu}(x) ,
\eqno(1.4)
$$
where $d^2\!\sigma^{\mu\nu}$ is the usual surface element,
$$
F_{\mu\nu}(x) = \d_{\mu}A_\nu(x) - \d_{\nu}A_\mu(x) ,
\eqno(1.5)
$$
that is, $F=\dd A$, and the greek indices $\mu, \nu,\dots$ take
values from $1$ to $D$.  If $D=3$, as in elementary vector calculus,
the right hand side of \eq{1.4} can be written as
$$
\int\limits_{\S} d^2\!\sigma^{\mu\nu} F_{\mu\nu}(x) =
       \int\limits_{\S} d^2\!\tl \sigma_\mu \tl F^\mu ,
$$
which leads to the familiar relation between the circulation of the
covariant vector field $A$ (eg. the vector potential or the magnetic
field) and the flux of the contravariant vector field $\tl F$
(correspondingly, the magnetic field or the current density).

We can restate our earlier remarks:  $\G[q]$ in \eq{1.4} is
invariant under the Abelian gauge transformations
$$
A_\mu \to A'_\mu = A_\mu + \d_\mu \Lambda ,
\eqno(1.6)
$$
and depends on the curve $q$ but not on the surface $\S$.  As shown
in \eq{1.2}, this is a consequence of $\dd F = \dd\,\dd A =0$ or,
in terms of the dual field
$$
\tl F_{\mu_1\cdots\mu_{\!D-2}} = {1 \over 2!} \,
    \ep_{\mu_1\cdots\mu_{\!D-2}\nu_1\nu_2} F^{\nu_1\nu_2} ,
$$
is a consequence of Bianchi's identity
$$
\d_{\mu_1} \tl F^{\mu_1\cdots\mu_{\!D-2}} = 0 .
\eqno(1.7)
$$

\St{} is used ocasionally in situations involving a flux through
a physical surface.  However, very often the surface $\S$ is an
auxiliary object, while the importance of the theorem stems from
the interest in exchanging the ``gauge potential'' $A_\mu$ in
favor of its gauge invariant ``field strength'' $F_{\mu\nu}$.
{}From this point of view, \St{} as formulated in \eq{1.4} is very
limited.  It requires the use of a fiducial surface $\S$, with
no immediate physical significance, and more important, it's
application is limited to those loops which are boundaries of
orientable surfaces.  The gauge invariant left hand side of \eq{1.4}
is well defined for a very wide class of loops, including self
intersecting and knotted loops (with the appropriate limitations
arising from the singularities of $A_\mu$).  But \St, as formulated
in \eq{1.4}, cannot be used to obtain an expression in terms of the
manifestly gauge invariant field strength $F_{\mu\nu}$, except for
the simplest loops.  This limitation is of increasing practical
importance in view of the growing role of the so-called Loop Space
variables in the analysis of gauge theories in general
\cite{\RevLoops}, of quantum gravity \cite{\RevGrav}, of topological
field theories \cite{\RevTFT}, and more recently in such classical
areas as the theory of turbulence \cite{\RevTurb}.  In many
applications the Abelian invariance, as expressed in eqs.~(1.3)
and~(1.6), is also a limitation.  A non-Abelian extension has been
discussed elsewhere \cite{\Bralic,\OtherNonAbelianStokes}, but we
shall not consider it here.

For Abelian gauge fields one can always fix the gauge completely
and then invert \eq{1.5} to obtain $A_\mu$ as a functional of
$F_{\mu\nu}$.  Thus, it is clear that in principle one can always
write the line integral in the left hand side of \eq{1.4} in terms
of the gauge invariant field strength $F_{\mu\nu}$.  The question
is then whether one can derive a simple geometrical expression
providing a direct generalization of \St{} in \eq{1.4} for general
loops in arbitary dimension $D$.  From the preceeding discussion it
should be clear that such a generalization must not involve any
auxiliary surface, so that whatever limitations apply follow solely
from the loop $q$ and the gauge potential $A_\mu$.  In the
following sections we derive and discuss various aspects of that
generalization of \St.
\varvskip{2}
\section{2. Vector and Tensor Currents}
\noindent
Consider a closed loop $q$ embedded in a $D$-dimensional manifold
$\M$ which we shall assume to be flat and, for simplicity, we take
to be Euclidean.  The loop $q$ will be parametrized by a parameter
$s \in [0,1]$, so it is given by a trajectory $q^\mu(s)$ in $\M$,
with $q^\mu(0) = q^\mu(1)$.  Associated to the loop there is a
vector current given by
$$
j^\mu [q;x] = \int_0^1 ds\,\dot q^\mu(s)\,\delta(x-q(s))
\eqno(2.1)
$$
where $\delta(x)$ is the $D$-dimensional delta in $\M$, and
$\dot q^\mu(s) = dq^\mu(s)/ds$.  Then, the loop integral in the
left hand side of \eq{1.4} is
$$
\G[q]
  = \int_0^1 ds\,\dot q^\mu(s)\,A_\mu(q(s))
  = \int d^D\!x \, j^\mu[q;x] \, A_\mu(x) ,
\eqno(2.2)
$$
and its invariance under the gauge transformation in \eq{1.6} is a
consequence of $j^\mu$ being conserved
$$
\d_\mu j^\mu[q;x] = 0,
\eqno(2.3)
$$
which in turn follows from the definition of the current in
\eq{2.1} and the fact that $q$ is a closed loop.

A key ingredient in the generlization of \St{} is the fact that
the current $j^\mu$ is itself the divergence of an antisymmetric
tensor current,
$$
j^\mu[q;x] = \d_\nu \tl\th^{\mu\nu}[q;x]
\eqno(2.4)
$$
with
$$
\tl\th^{\mu\nu} = - \tl\th^{\nu\mu} .
$$
Indeed, consider
$$
\tl\th_{\mu\nu}[q;x] = - \int_0^1 ds\,
[ \dot q_\mu(s)\,\d_\nu - \dot q_\nu(s)\,\d_\mu ]
\Gd(x-q(s)) ,
\eqno(2.5)
$$
were $\Gd(x)$ is the inverse of the laplacian in $D$-dimensions
$$
-\d^2 \Gd(x) = \delta(x) .
\eqno(2.6)
$$
The divergence of $\tl\th^{\mu\nu}$ is
$$
\d_\nu \tl\th^{\mu\nu} [q;x] = \int_0^1 ds\,
[ \dot q^\nu(s)\d_\nu\d^\mu - q^\mu(s) \d^2 ] \Gd(x-q(s))
$$
The first term in the right hand side vanishes for closed loops
$$
\int_0^1 ds \, \dot q^\nu(s)\d_\nu\d_\mu \Gd(x-q(s)) =
- \int_0^1 ds \,{d \over ds} \d_\mu \Gd(x-q(s)) = 0 ,
$$
so \eq{2.4} follows after using \eq{2.6}.  Then, substituting
\eq{2.4} into \eq{2.2} we have
$$
\eqalign{
\int d^D\!x \, j^\mu[q;x] \, A_\mu(x)
 &= \int d^D\!x \, \d_\nu \tl\th^{\mu\nu}[q;x] \, A_\mu(x) \cr
 &= -\int d^D\!x \, \tl\th^{\mu\nu}[q;x] \, \d_\nu A_\mu(x) .\cr
}
$$
Thus,
$$
\eqalign{
\int_0^1 ds\,\dot q^\mu(s)\,A_\mu(q(s))
&= \int d^D\!x \, j^\mu[q;x] \, A_\mu(x) \cr
&= {1 \over 2} \int d^D\!x \, \tl\th^{\mu\nu}[q;x]
\, F_{\mu\nu}(x) , \cr
}
\eqno(2.7)
$$
which is the generalized version of \St{} we were after.
\varvskip{2}
\section{3. The two-dimensional case}
\noindent
To get a feeling for the contents of this result let us consider
first the $D=2$ case, when \eq{2.7} can be written as
$$
\int_0^1 ds\,\dot q^\mu(s)\,A_\mu(q(s)) =
\int d^2\!x \,{\th[q;x]}\,\tl F(x) ,
\eqno(3.1)
$$
where
$$
\tl F = {1 \over 2!} \,\ep^{\mu\nu} F_{\mu\nu}
$$
and
$$
\th[q;x] = {1 \over 2!} \,\ep_{\mu\nu} \, \tl\th^{\mu\nu}[q;x] .
\eqno(3.2)
$$
Substituting here \eq{2.5}, we get
$$
\th[q;x] = \int_0^1 ds\,\dot q^\mu(s)\,\B_\mu(x-q(s)) ,
\eqno(3.3)
$$
where $\B_\mu$ can be chosen to be
$$
\B_\mu(x) = - \ep_{\mu\nu} \, \d^\nu G_{\!2}(x) .
\eqno(3.4)
$$
For our discussion in higher dimensions it will be relevant to
note here that this choice for $\B_\mu$ is not unique.  Indeed,
\eq{3.3} defines $\B_\mu$ only up to a gauge transformation
$$
\B_\mu(x) \to \B_\mu(x) + \d_\mu \Phi(x) ,
\eqno(3.5)
$$
under which $\th$ and $\tl\th^{\mu\nu}$, and consequently $j^\mu$,
remain invariant.

To proceed, recall that in $D=2$ the inverse laplacian can be chosen
to be
$$
G_{\!2}(x) = - {1 \over 4\pi} \, \ln m^2 x^2 ,
\eqno(3.6)
$$
so $\B_\mu$ in \eq{3.4} becomes
$$
\B_\mu(x) = {1 \over 2\pi} \,\ep_{\mu\nu} \, {x^\nu \over x^2} .
\eqno(3.7)
$$
Thus, $\B_\mu(x)$ can be thought of as the ``gauge potential'' at
the point $x$ due to a (anti-) monopole at the origin, so $\th[q;x]$
in \eq{3.3}, and therefore $\tl\th^{\mu\nu}[q;x]$, is the winding
number of the loop $q$ around the point $x$.  Then \eq{3.1} is the
usual \St{} in $D=2$, except that the factor $\th[q;x]$ in the
integrand provides the correct weight to each area element,
rendering the theorem valid for arbitrary loops, even if they are
self intersecting or self overlapping~\cite{\Bralic}.
\varvskip{2}
\section{4. Hidden gauge invariance of higher rank}
\noindent
In higher dimensions the situation is more involved.  We can think
of the right hand side of \eq{2.7} as a sum of projections onto
the individual $\mu$--$\nu$ planes, with $\tl\th^{\mu\nu}[q;x]$
providing the correct weight to the surface element in each integral.
But for $D>2$ there is more structure than that in this generalized
\St{}.  Although we have disposed of the surface used in the
usual formulation of the theorem, we can expect that the symmetry
under deformations of that surface may remain hidden somewhere.

Analogous to $\th[q;x]$ in \eq{3.2}, for $D>2$ we have
$$
\th_{\mu_1\cdots\mu_{\!D-2}}[q;x] = {1 \over 2!} \,
\ep_{\mu_1\cdots\mu_{\!D-2}\nu_1\nu_2} \,
\tl\th^{\nu_1\nu_2}[q;x]
\eqno(4.1)
$$
and conversely,
$$
\tl\th_{\mu\nu}[q;x] = {1 \over (D-2)!} \,
\ep_{\mu\nu\lambda_1\cdots\lambda_{\!D-2}} \,
\th^{\lambda_1\cdots\lambda_{\!D-2}}[q;x] .
\eqno(4.2)
$$
Then, as in \eq{3.3}, we now have
$$
\th_{\mu_1\cdots\mu_{\!D-2}}[q;x] =
\int_0^1 ds\,\dot q^\nu(s)\,\B_{\mu_1\cdots\mu_{\!D-2}\nu}(x-q(s)) ,
\eqno(4.3)
$$
where we can choose
$$
\B_{\mu_1\cdots\mu_{\!D-1}}(x) =
- \ep_{\mu_1\cdots\mu_{\!D-1}\lambda} \, \d^\lambda \Gd(x) .
\eqno(4.4)
$$

As with $\B_\mu$ in \eq{3.4}, this choice for $\B$ is not unique.
Let us show that the invariance under the gauge transformation of
$\B$, given in $D=2$ by \eq{3.5}, now generalizes to
$$
\B_{\mu_1\cdots\mu_{\!D-1}}(x) \to
\B_{\mu_1\cdots\mu_{\!D-1}}(x) +
(D-1) \, \d_{[\mu_1} \Phi_{\mu_2\cdots\mu_{\!D-1}]}(x) ,
\eqno(4.5)
$$
where
$\Phi_{\mu_1\cdots\mu_{\!D-2}} = \Phi_{[\mu_1\cdots\mu_{\!D-2}]}$ is
a completely antisymmetric tensor field.  Substituting into \eq{4.3}
one finds that under this transformation $\th$ transforms as
$$
\th_{\mu_1\cdots\mu_{\!D-2}}[q,x] \to
\th_{\mu_1\cdots\mu_{\!D-2}}[q,x] +
(D-2) \, \d_{[\mu_1} \Psi_{\mu_2\cdots\mu_{\!D-2}]}[q,x]
\eqno(4.6)
$$
where
$$
\Psi_{\mu_1\cdots\mu_{\!D-3}}[q,x] =
\int_0^1 ds\,\dot q^\nu(s)\,\Phi_{\mu_1\cdots\mu_{\!D-3}\nu}(x-q(s)) .
\eqno(4.7)
$$
Correspondingly, for $\tl\th$ we get, with \eq{4.2},
$$
\tl\th^{\mu\nu}[q,x] \to
\tl\th^{\mu\nu}[q,x] + {1 \over (D-3)!} \,
\ep^{\mu\nu\lambda_1\cdots\lambda_{\!D-2}}
\d_{\lambda_1} \Psi_{\lambda_2\cdots\lambda_{\!D-2}}[q,x] .
\eqno(4.8)
$$
Hence, in $D>2$ neither $\th$ nor $\tl\th$ are invariant, but
the divergence of $\tl\th$ is invariant
$$
\d_\nu \tl\th^{\mu\nu}[q;x] \to
\d_\nu \tl\th^{\mu\nu}[q;x] .
$$

Then, in order to have $j^\mu = \d_\nu \tl\th^{\mu\nu}$ invariant,
as needed by \St{} in \eq{2.7}, $\th[q,x]$ is the line integral of
a gauge potential $\B(x)$ of rank $(D-1)$ (in the sense of
Kalb-Ramond-Nambu~\cite{\KalbRamondNambu}).  Moreover, $\th[q,x]$
itself transforms as a gauge potential of rank $(D-2)$, but the
corresponding gauge transformation depends on the loop $q$.  The
sources for these potentials are different: \eq{4.4} shows that
$\B(x)$ has as source an ordinary point singulariry, while the
source for $\th[q,x]$ has a singularity along the loop $q$ as
seen from \eq{4.3}.  In $D=2$ the potential $\B$ becomes of rank
$1$, while $\th$ is invariant.

Finally, let us clarify the relation between this higher rank gauge
invariance in $D>2$, and the invariance under deformations of the
auxiliary surface in the usual formulation of \St, also in $D>2$.
The generalized \St{} in \eq{2.7} is
$$
\eqalign{
\G[q]
&= {1 \over 2} \int dx \, \tl\th^{\mu\nu}[q;x] \, F_{\mu\nu}(x) \cr
&= {1 \over (D-2)!}\int dx \,
\th_{\mu_1\cdots\mu_{\!D-2}}[q;x] \,
\tl F^{\mu_1\cdots\mu_{\!D-2}}(x) . \cr
}
$$
Then, under the transformation of $\th$ in \eq{4.6}, $\G$ would
change by
$$
\eqalign{
\delta \G
&= {1 \over (D-3)!} \int dx \,
   \d_{\mu_1} \Psi_{\mu_2\cdots\mu_{\!D-2}}[q,x]
   \, \tl F^{\mu_1\cdots\mu_{\!D-2}}(x) \cr
&= -{1 \over (D-3)!} \int dx \,
    \Psi_{\mu_2\cdots\mu_{\!D-2}}[q,x]
    \, \d_{\mu_1} \tl F^{\mu_1\cdots\mu_{\!D-2}}(x) , \cr
}
$$
which vanishes by virtue of Bianchi's identity in \eq{1.7}.  Thus,
the higher rank gauge invariance exhibited in $D>2$ has the same
geometrical contents as the invariance under deformations of the
auxiliary surface in ordinary \St.  It may be noticed that in
$D=2$ there is neither Bianchi identity, nor room where to deform
the surface.
\varvskip{2}
\section{Acknowledgments}
\noindent
This work was partially supported by FONDECYT, under grant 751/92,
and by Fundaci\'on Andes.
\varvskip{2}
\section{References}
\ref{\RevLoops}
  {For a review see R. Loll, {\sl Chromodynamics and gravity as
  theories on loop space}, Pennsylvania State University preprint
  CGPG-93/9-1, September 1993.}
\ref{\RevGrav}
  {C. Rovelli and L. Smolin, \NP{B331} (1990) 80.}
\ref{\RevTFT}
  {For reviews see M. Blau and G. Thompson, Lectures at the 1993
  Trieste Summer School on High Energy Physics and Cosmology,
  Preprint IC/93/356, October 1993;  E. Guadagnini, \IJMP{A7}
  (1992) 877.}
\ref{\RevTurb}
  {A. A. Migdal, {\sl Loop equation and area law in turbulence},
  Princeton University preprint PUPT-1246, October 1993.}
\ref{\Bralic}
  {N. Brali\'c, \PR{D22} (1980) 3090.}
\ref{\OtherNonAbelianStokes}
  {I. Ya. Aref'eva, {\sl Theor.\ Math.\ Phys.\ }{\bf 43} (1980) 353;
  P. M. Fishbane, S. Gasiorowicz and P. Kaus, \PR{D24} (1981) 2324;
  L. Gross, {\sl J.\ Funct.\ Anal.\ }{\bf 63} (1985) 1.}
\ref{\KalbRamondNambu}
  {M. Kalb and P. Ramond, \PR{D9} (1974) 2273; Y. Nambu, \PRp{23}
  (1976) 250.}
\bye